\documentclass[12pt]{article}
\usepackage{color}
\usepackage{graphicx}
\def\nabstar#1{\nabla\kern-0.5pt\smash{\raise 4.5pt\hbox{$\ast$}}
               \kern-4.5pt_{#1}}

\def\drvstar#1{\partial\kern-0.5pt\smash{\raise 4.5pt\hbox{$\ast$}}
               \kern-5.0pt_{#1}}

\def\newline{\relax\ifhmode\null\hfil\break\else\nonhmodeerr@\newline\fi}
\def\frac#1#2{{#1\over#2}}
\def\text#1{{\hbox{\rm #1}}}
\def\flushpar{{\par \noindent}}

\newcommand{\beq}{\begin{equation}}
\newcommand{\eeq}{\end{equation}}
\newcommand{\bea}{\begin{eqnarray}}
\newcommand{\eea}{\end{eqnarray}}

\def\EQ{\hspace{-2mm} &=& \hspace{-2mm}}

\def\BA{\begin{eqnarray}}
\def\EA{\end{eqnarray}}
\def\BAN{\begin{eqnarray*}}
\def\EAN{\end{eqnarray*}}

\def\g5{\gamma_5}
\def\g4{\gamma_4}
\def\g3{\gamma_3}
\def\g2{\gamma_2}
\def\g1{\gamma_1}
\def\gi{\gamma_i}

\def\u{{\bf u}}
\def\d{{\bf d}}
\def\s{{\bf s}}
\def\c{{\bf c}}
\def\q{{\bf q}}
\def\Q{{\bf Q}}
\def\ubar{\bar{\bf u}}
\def\dbar{\bar{\bf d}}
\def\sbar{\bar{\bf s}}
\def\cbar{\bar{\bf c}}
\def\qbar{\bar{\bf q}}

\input epsf.sty
\newdimen\psfigsize
\def\psfigure#1 #2 #3 #4 #5{
    \begin{figure}[tbh]
      \begin{center}
      \vbox{
        \null\vskip-0.2in\hskip#2
        \epsfxsize=#1
        \epsfbox{#4}
        \vskip -0.3in
        \caption {#5 \label{#3}}
        \vskip 0.0 true in plus 0.3 true in
      }
      \end{center}
   \end{figure}
}
\usepackage{graphicx}
\begin{document}
\thispagestyle{empty}
\begin{flushright}
NTUTH-06-505B \\
April 2006 \\
\end{flushright}
\vskip 2.5truecm
\begin{center}
{\LARGE Pseudovector meson with strangeness and closed-charm}
\end{center}
\vskip 1.0truecm
\centerline{{\bf Ting-Wai~Chiu$^{1}$, Tung-Han~Hsieh$^{2}$}}
\vskip2.0ex
\centerline{$^1\hskip-3pt$ \it
Department of Physics, National Taiwan University,}
\vskip1.0ex
\centerline{\it Taipei, 10617, Taiwan}
\vskip2.0ex
\centerline{$^2\hskip-3pt$ \it
Physics Section, Commission of General Education,}
\vskip1.0ex
\centerline{\it National United University, Miao-Li, 36003, Taiwan}
\vskip1.0ex
\vskip2.0ex
\centerline{\bf (TWQCD Collaboration)}
\vskip 1cm
\bigskip \nopagebreak \begin{abstract}

\noindent

We investigate the mass spectrum of $ 1^{+} $ exotic 
mesons with quark content $ (\c\s\cbar\qbar) $/$ (\c\q\cbar\sbar) $,   
using molecular and diquark-antidiquark operators,   
in quenched lattice QCD with exact chiral symmetry.
For the molecular operator 
$ \{ (\qbar\gamma_i\c)(\cbar\gamma_5\s)-
     (\cbar\gamma_i\s)(\qbar\gamma_5\c) \} $ 
and the diquark-antidiquark operator 
$ \{ (\q^T C \gi \c )(\sbar C \gamma_5 \cbar^T) 
    -(\q^T C \gamma_5 \c)(\sbar C \gi^T \cbar^T) \} $,  
both detect a $ 1^{+} $ resonance with mass around $ 4010 \pm 50 $ MeV
in the limit $ m_q \to m_{u,d} $.


Charmed Mesons, Diquarks

\end{abstract}
\vskip 1.5cm 
\newpage\setcounter{page}1

\section{Introduction}

Since the discovery of $ D_s(2317) $ \cite{Aubert:2003fg} by BABAR 
in April 2003, a series of new heavy mesons\footnote{ 
For recent reviews of these new heavy mesons,  
see, for example, 
Ref. \cite{Swanson:2006st}, 
and references therein.} 
with open-charm and closed-charm have been observed by 
Belle, CDF, CLEO, BABAR, and BES. 
Among these new heavy mesons, the most intriguing ones are 
the charmonium-like states, 
$ X(3872) $ \cite{Choi:2003ue}, 
$ Y(3940) $ \cite{Abe:2004zs}, 
$ Y(4260) $ \cite{Aubert:2005rm}, 
$ Z(3930) $ \cite{Uehara:2005qd},
and $ X(3940) $ \cite{Abe:2005hd}. 
Evidently, one can hardly interpret all of them as orbital and/or radial 
excitations in the charmonium spectrum. 
Thus it is likely that some of them are exotic (non-$q\bar q $) mesons 
(e.g., molecule, diquark-antidiquark, and hybrid meson).
Theoretically, the central question is whether  
the spectrum of QCD possesses these resonances, with the correct
quantum numbers, masses, and decay widths.  

Recently, we have investigated the mass spectrum of closed-charm 
exotic mesons with $ J^{PC} = 1^{--} $ 
\cite{Chiu:2005ey}, and $ 1^{++} $ \cite{Chiu:2006hd}, 
in lattice QCD with exact chiral symmetry 
\cite{Kaplan:1992bt,Narayanan:1995gw,Neuberger:1997fp,
      Ginsparg:1981bj,Chiu:2002ir}. 
By constructing molecular and diquark-antidiquark operators
with quark content $ (\c\q\cbar\qbar) $, 
we measured their time-correlation functions, 
and extracted the mass spectrum of the hadronic states overlapping 
with these exotic meson operators.   
Our results suggest that $ Y(4260) $ and $ X(3872) $
are in the spectrum of QCD, 
with $ J^{PC} = 1^{--} $ and $ 1^{++} $ respectively,  
and both with quark content $ (\c\u\cbar\ubar) $. Note that we have 
been working in the isospin limit (with $ m_u = m_d $), thus our 
results \cite{Chiu:2005ey,Chiu:2006hd} also imply the existence
of exotic mesons with quark content $ (\c\d\cbar\dbar) $, even though 
we cannot determine their mass differences from those with 
$ (\c\u\cbar\ubar) $. 
Moreover, we also observe heavier exotic mesons 
with quark contents $ (\c\s\cbar\sbar) $ and $ (\c\c\cbar\cbar) $, 
for $ J^{PC} = 1^{++} $ \cite{Chiu:2006hd}, and $ 1^{--} $   
\cite{Chiu:2005ey}.  

Now if the spectrum of QCD does possess exotic mesons with quark 
content $ (\c\q\cbar\qbar) $, then it is likely that there
are also exotic mesons with other quark contents, 
e.g., $ (\c\q\cbar\sbar) $ and $ (\c\s\cbar\qbar) $.  
However, in general, whether any combination of two quarks 
and two antiquarks can emerge as a hadronic state relies on  
the nonperturbative dynamics between these four quarks.    

In this paper, we investigate the masses of the lowest-lying 
states (with $ J^P = 1^+ $) of molecular and diquark-antidiquark operators 
with quark content $ (\c\s\cbar\qbar) $ or $ (\c\q\cbar\sbar) $. 
For two lattice volumes $ 24^3 \times 48 $ and $ 20^3 \times 40 $, 
each of 100 gauge configurations generated with single-plaquette action 
at $ \beta = 6.1 $, we compute point-to-point quark propagators   
for 30 quark masses in the range $ 0.03 \le m_q a \le 0.80 $, 
and measure the time-correlation functions of these exotic meson operators.  
The inverse lattice spacing $ a^{-1} $
is determined with the experimental input of $ f_\pi $. 
The strange quark bare mass $ m_s a = 0.08 $, and the charm
quark bare mass $ m_c a = 0.80 $ are fixed such that the masses of
the corresponding vector mesons are in good agreement with  
$ \phi(1020) $ and $ J/\psi(3097) $ respectively 
\cite{Chiu:2005zc,Chiu:2005ue}.

\section{The Molecular Operator}

In general, one can construct many different molecular operators with  
quark content $(\c\q\cbar\sbar)$ or $ (\c\s\cbar\qbar) $ 
such that the lowest-lying state of 
each operator has $ J^{P} = 1^{+} $. However, not every one of them has
good overlap with the lowest-lying hadronic state having the same 
quark content and quantum numbers.
In the following we only present one of the good molecular operators 
\footnote{The operators we have investigated include  
$ (\qbar\gi\c)(\cbar\gamma_5\s) $, 
$ (\cbar\gi\s)(\qbar\gamma_5\c) $, 
and $ [(\qbar\gi\c)(\cbar\gamma_5\s)\pm(\cbar\gi\s)(\qbar\gamma_5\c)] $, 
and they all yield compatible masses for the lowest-lying meson.}.
Explicitly,     
\bea
\label{eq:DVDS5A}
M \EQ \frac{1}{\sqrt{2}} \left\{ (\qbar\gi\c)(\cbar\gamma_5\s)
                                -(\cbar\gi\s)(\qbar\gamma_5\c) \right\} 
\eea

\begin{figure}[htb]
\begin{center}
\includegraphics*[height=10cm,width=8cm]{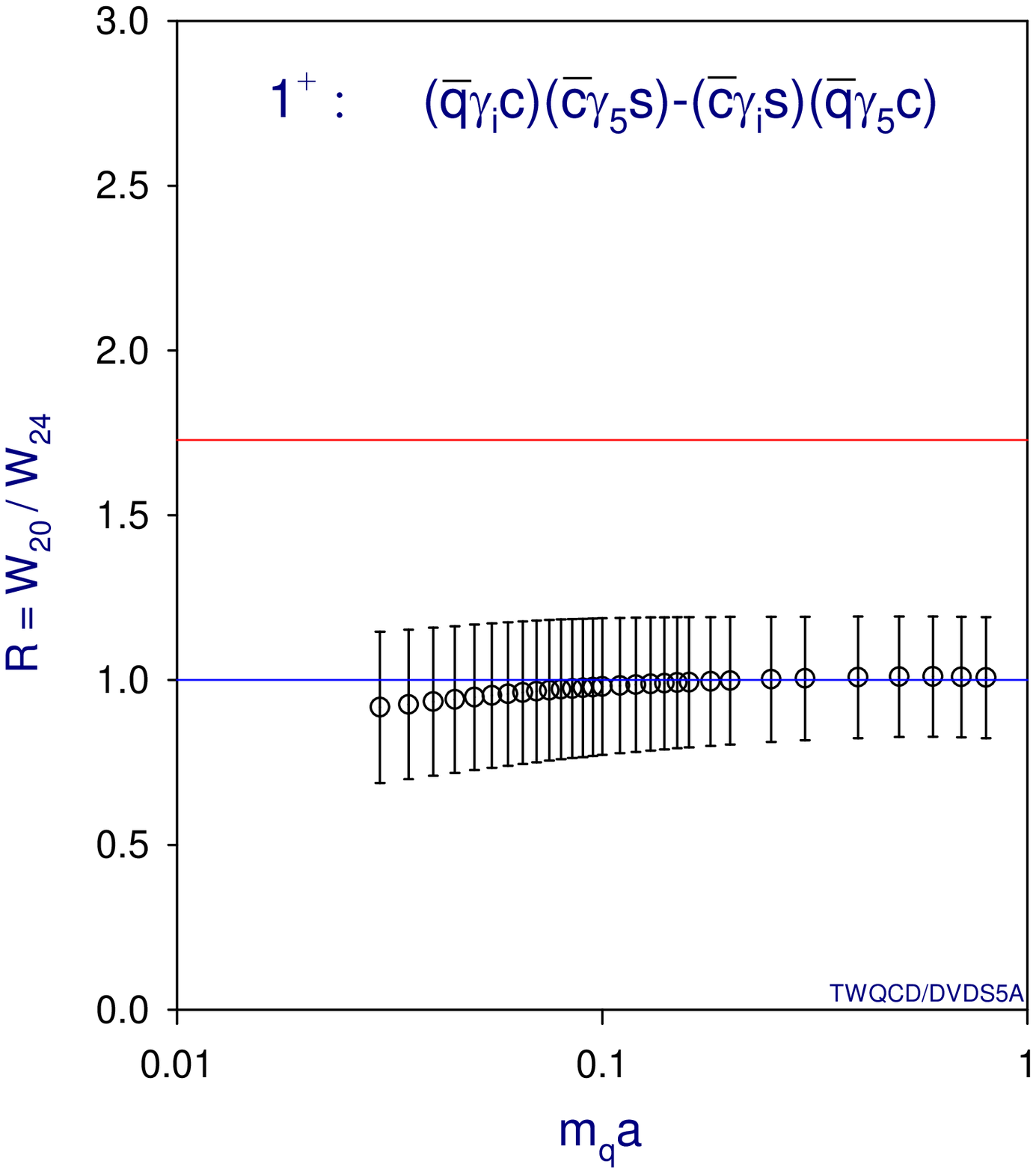}
\caption{
The ratio of spectral weights of the lowest-lying state
of the molecular operator $ M $,
for $ 20^3 \times 40 $ and $ 24^3 \times 48 $ lattices at $ \beta = 6.1 $.
The upper-horizontal line $ R = (24/20)^3 = 1.728 $,
is the signature of 2-particle scattering state,
while the lower-horizontal line $ R = 1.0 $ is the signature
of a resonance.}
\label{fig:sw2024_DVDS5A}
\end{center}
\end{figure}

\begin{figure}[htb]
\begin{center}
\includegraphics*[height=10cm,width=8cm]{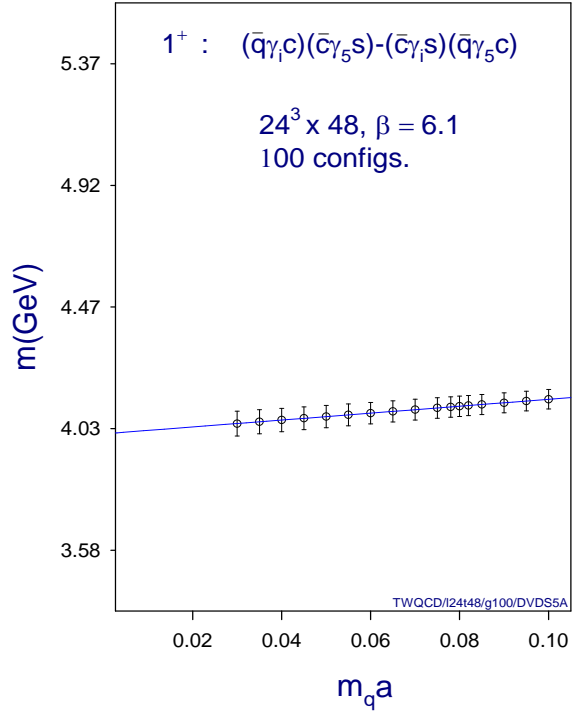}
\caption{
The mass of the lowest-lying state
of $ M $ versus the quark mass $ m_q a $, 
on the $ 24^3 \times 48 $ lattice at $ \beta = 6.1 $.
The solid line is the linear fit.}
\label{fig:mass_DVDS5A}
\end{center}
\end{figure}

We measure the time-correlation function, 
\BAN
C(t)=\sum_{\vec{x}} \left< M(\vec{x},t) M^\dagger(\vec{0},0) \right>
\EAN
where the $ \c\cbar $ annihilation diagrams are neglected 
such that $ C(t) $ does not overlap with any conventional meson  
states. Also, $ C(t) $ is averaged over $ C_i $ (with $ \gamma_i $) 
for $ i=1,2,3 $, where in each case,
the ``forward-propagator" $ C_i(t) $ and ``backward-propagator"
$ C_i(T-t) $ are averaged to increase the statistics.
The same strategy is applied to all time-correlation functions
in this paper. Then the average of $ C(t) $ over all 
gauge configurations is fitted to the usual formula 
\BAN
\frac{Z}{2 m a } [ e^{-m a t} + e^{-m a (T-t)} ]
\EAN
to extract the mass $ m a $ of the lowest-lying state 
and its spectral weight
\BAN
W = \frac{Z}{2 m a } \ .
\EAN
Theoretically, if this state is a genuine resonance, then its mass $ m a $  
and spectral weight $ W $ should be almost constant for 
any lattices with the same lattice spacing. On the 
other hand, if it is a 2-particle scattering state, then its mass 
$ m a $ is sensitive to the lattice volume, and its spectral
weight is inversely proportional to the spatial volume for lattices
with the same lattice spacing.
In the following, we shall use the ratio of the spectral weights on 
two spatial volumes $ 20^3 $ and $ 24^3 $ with the same lattice spacing 
($\beta = 6.1 $) to discriminate whether any hadronic state under 
investigation is a resonance or not.

In Fig. \ref{fig:sw2024_DVDS5A}, the ratio ($ R=W_{20}/W_{24} $) 
of spectral weights of the lowest-lying state extracted from 
the time-correlation function of $ M $ on the 
$ 20^3 \times 40 $ and $ 24^3 \times 48 $ lattices is plotted 
versus the quark mass $ m_q a \in [0.03, 0.80] $.   
(Here $ \q $ ($\qbar$) is always taken to be different from 
$\c$ ($\cbar$) and $\s$ ($\sbar$), 
even in the limit $ m_q \to m_c $ or $ m_s $).
Evidently, $ R \simeq 1.0 $ for the entire range of quark masses,   
which implies that there exist $ J^{P} = 1^{+} $ resonances, 
with quark contents $ (\c\s\cbar\ubar) $ and $ (\c\s\cbar\dbar) $ 
respectively.

In Fig. \ref{fig:mass_DVDS5A}, the mass of the lowest-lying state 
extracted from the molecular operator $ M $ is plotted versus $ m_q a $.  
In the limit $ m_q \to m_{u,d} \simeq 0.00265 a^{-1} $ 
(corresponding to $ m_\pi = 135 $ MeV), it gives $ m = 4007(34) $ MeV.

\section{The Diquark-Antidiquark Operator}

\begin{figure}[htb]
\begin{center}
\includegraphics*[height=10cm,width=8cm]{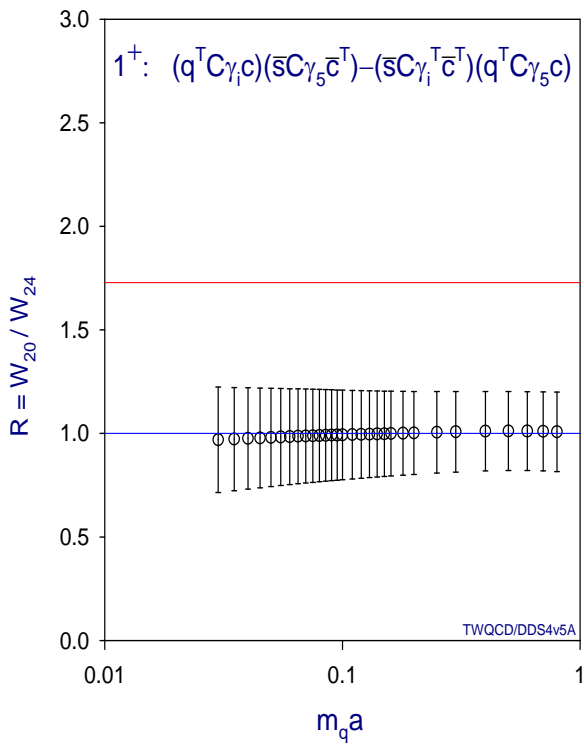}
\caption{
The ratio of spectral weights of the lowest-lying state
of diquark-antidiquark operator $ D_4 $, 
for $ 20^3 \times 40 $ and $ 24^3 \times 48 $ lattices at $ \beta = 6.1 $.
The upper-horizontal line $ R = (24/20)^3 = 1.728 $,
is the signature of 2-particle scattering state,
while the lower-horizontal line $ R = 1.0 $ is the signature
of a resonance.}
\label{fig:sw2024_DDS4V5A}
\end{center}
\end{figure}

\begin{figure}[htb]
\begin{center}
\includegraphics*[height=10cm,width=8cm]{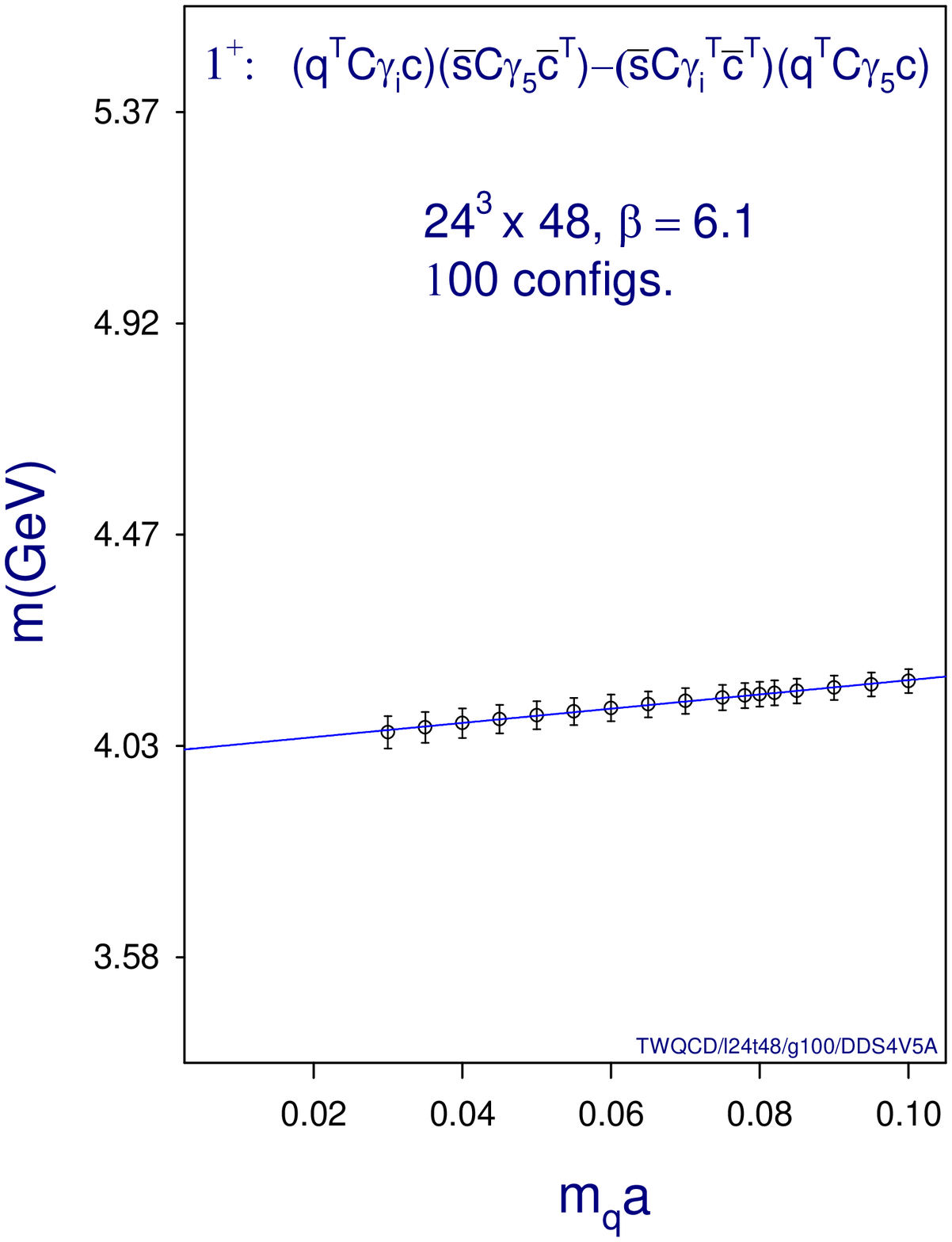}
\caption{
The mass of the lowest-lying state of the diquark-antidiquark operator
$ D_4 $ versus the quark mass $ m_q a $,
on the $ 24^3 \times 48 $ lattice at $ \beta = 6.1 $.
The solid line is the linear fit.}
\label{fig:mass_DDS4V5A}
\end{center}
\end{figure}

In general, one can construct many different diquark-antiquark operators 
with quark content $(\c\q\cbar\sbar)$ or $(\c\s\cbar\qbar)$ 
such that the lowest-lying 
state of each operator has $ J^{P} = 1^{+} $. However, not every one
of them has good overlap with the lowest-lying hadronic state having 
the same quark content and quantum numbers. 
In the following we only present one of the good diquark-antidiquark 
operators. Explicitly,     
\bea
\label{eq:DDS4V5A}
D_4(x)
\EQ \frac{1}{\sqrt{2}} \left\{ 
   (\q^T C \gi \c )_{xa} (\sbar C \gamma_5 \cbar^T)_{xa} 
  -(\sbar C \gi^T \cbar^T)_{xa} (\q^T C \gamma_5 \c)_{xa} \right\} 
\eea
where $ C $ is the charge conjugation operator satisfying
$ C \gamma_\mu C^{-1} = -\gamma_\mu^T $ and
$ (C \gamma_5)^T=-C\gamma_5 $. Here the 
``diquark" operator $ (\q^T \Gamma \Q)_{xa} $ for any Dirac matrix 
$ \Gamma $ is defined as
\bea
\label{eq:diquark}
({\q}^T \Gamma {\Q} )_{xa} \equiv \epsilon_{abc} 
 {\q}_{x\alpha b} \Gamma_{\alpha\beta} {\Q}_{x\beta c}
\eea
where 
$ x $, $ \{ a,b,c \} $ and $ \{ \alpha, \beta \} $
denote the lattice site, color, and Dirac indices respectively,  
and $ \epsilon_{abc} $ is the completely antisymmetric tensor. 
Thus the diquark (\ref{eq:diquark}) transforms like color anti-triplet.  
For $ \Gamma = C \gamma_5 $, it transforms like $ J^P = 0^{+} $,  
while for $ \Gamma = C \gamma_i $ ($i=1,2,3 $), it 
transforms like $ 1^{+} $. 

In Fig. \ref{fig:sw2024_DDS4V5A}, the ratio ($ R=W_{20}/W_{24} $)
of spectral weights of the lowest-lying state extracted from
the time-correlation function of $ D_4 $ on the $ 20^3 \times 40 $
and $ 24^3 \times 48 $ lattices is plotted
versus the quark mass $ m_q a \in [0.03, 0.8] $.
Evidently, $ R \simeq 1.0 $ for the entire range of quark masses,   
which implies that there exist $ J^{P} = 1^{+} $ resonances, 
with quark contents $ (\c\d\cbar\sbar) $ and $ (\c\u\cbar\sbar) $ 
respectively.

In Fig. \ref{fig:mass_DDS4V5A}, the mass of the lowest-lying state
of the diquark-antidiquark operator $ D_4 $
is plotted versus $ m_q a $. In the limit $ m_q \to m_{u,d} $, 
it gives $ m = 4015(25) $ MeV. 

\begin{table}
\begin{center}
\begin{tabular}{c|c|c}
Operator & Mass (MeV) & R/S  \\
\hline
$\frac{1}{\sqrt{2}}[ (\ubar \gamma_i \c)(\cbar \gamma_5 \s)
                    -(\cbar \gamma_i \s)(\ubar \gamma_5 \c) ] $
        &   4007(34)(31) &   R  \\
$ \frac{1}{\sqrt{2}}\left\{(\u^T C\gamma_i \c)(\sbar C\gamma_5\cbar^T)
  -(\u^T C \gamma_5 \c)(\sbar C\gamma_i^T \cbar^T) \right\} $      &
 4015(25)(27) & R  \\
\hline
\end{tabular}
\caption{Masses of the lowest-lying states ($ J^{P} = 1^{+} $) of the 
molecular operator $ M $ and the diquark-antidiquark operator $ D_4 $. 
The last column R/S denotes resonance (R) or scattering (S) state.}
\label{tab:mass_summary}
\end{center}
\end{table}

\section{Summary and Discussions}

In this paper, we have investigated the mass spectra of 
the molecular operator $ M $  
and the diquark-antidiquark operator $ D_4 $, 
in quenched lattice QCD with exact chiral symmetry. 
Our results for $ M $ and $ D_4 $ are summarized in 
Table \ref{tab:mass_summary}, 
where in each case, the first error is statistical, and 
the second one is our crude estimate of combined systematic uncertainty.

Evidently, both the molecular operator $ M $
and the diquark-antidiquark operator $ D_4 $ 
detect a $ 1^{+} $ resonance around $ 4010 \pm 50 $ MeV  
in the limit $ m_q \to m_{u,d} $. Since its mass is just 
slightly above Y(3940), high energy experiments should be able to 
see whether such a resonance, say, $ X_s $, exists in some decay channels,  
e.g., $ X_s \to K \pi J/\Psi $, in the near future.

Note that in our calculations, we have omitted 
all internal quark loops generated by vacuum fluctuations
(i.e., the quenched approximation). 
However, we expect that the quenched approximation
only results a few percent systematic error in the mass spectrum,
especially for charmed mesons \cite{Chiu:2005ue}.
If it turns out that high energy experiments 
do not find any charmonium-like resonances around 3960-4060 MeV, 
in particular, in the $ K \pi J/\Psi $ channel, then it might imply that  
quenched QCD could be dramatically different from the unquenched QCD, 
even for the mass spectrum.

\bigskip

\flushpar
{\bf Acknowledgement}

\noindent

\bigskip

We thank Steve Olsen for his question regarding $ (\c\u\cbar\sbar) $, 
which motivated our present study.   
This work was supported in part by the National Science Council,
Republic of China, under the Grant No. NSC94-2112-M002-016 (T.W.C.),  
and Grant No. NSC94-2119-M239-001 (T.H.H.), and by  
the National Center for High Performance Computation at Hsinchu, 
and the Computer Center at National Taiwan University.

\bigskip
\bigskip

\vfill\eject

\end{document}